\documentclass[english,aps,prl,10pt,twocolumn,showpacs,preprintnumbers]{revtex4-1}
\usepackage{amssymb}
\usepackage[T1]{fontenc}
\usepackage{latexsym,epsfig}
\usepackage{xcolor}
\usepackage[latin9]{inputenc}
\usepackage{graphicx}
\usepackage{amsfonts}
\usepackage{epsfig}
\usepackage{dcolumn}
\usepackage{bm}
\usepackage{babel}
\usepackage{hyperref}
\usepackage{amsmath,mathrsfs}
\usepackage{lipsum}

\makeatletter

\@ifundefined{textcolor}{}
{
 \definecolor{BLACK}{gray}{0}
 \definecolor{WHITE}{gray}{1}
 \definecolor{RED}{rgb}{1,0,0}
 \definecolor{GREEN}{rgb}{0,1,0}
 \definecolor{BLUE}{rgb}{0,0,1}
 \definecolor{CYAN}{cmyk}{1,0,0,0}
 \definecolor{MAGENTA}{cmyk}{0,1,0,0}
 \definecolor{YELLOW}{cmyk}{0,0,1,0}
 }

\makeatother

\begin{document}

\date{\today }
\title{Metastability on the steady states in a Fermi-like model of
counterflowing particles}
\author{Eduardo Velasco Stock, Roberto da Silva}
\affiliation{Instituto de F\'{\i}sica, Universidade Federal do Rio Grande do Sul, 
UFRGS, Porto Alegre - RS, 91501-970, Brasil.}

\begin{abstract}
In this work we propose a two-dimensional extension of a previously defined
one-dimensional version of a model of counterflowing particles, which
considers an adapted Fermi-Dirac distribution to describe the transition
probabilities. In this modified and extended version of the model, we
consider that only particles of different species interact and they hop
through the cells of a two-dimensional rectangular lattice with
probabilities taking into account diffusive and scattering aspects. We show
that for a sufficiently low level of randomness ($\alpha \geq 10$), the
system can relax to a mobile self-organized steady state of counterflow
(lane formation) or to an immobile state (clog) depending sensitively on the
initial conditions if the system has an average density near the crossover
value ($\rho _{c}$). We also show that for certain suitable mixing of the
species, we peculiarly have 3 different situations: (i) The immobile,(ii)
Mobile organized by lanes, and (iii) Mobile without lane formation for the
same density value. All of our results were obtained by performing Monte
Carlo simulations.
\end{abstract}

\maketitle

\label{Section:Introduction}

Counterflowing streams of particles \cite%
{Schmittmann-1992,Roberto-pedestrian-dynamics-2015,Andre-keep-left-behavior-2016,mestrado-eduardo-roberto2017}
can lead to interesting patterns such for example charged colloids \cite%
{Vissers-band-formation-2011}. Exactly as pedestrian organize their motion
to make that system flow, particles without a \textquotedblleft natural
intelligence\textquotedblright , self-propelled or simply oriented by a
field, sometimes can organize their motion by lanes. However, exclusion
effects seems to bring clogging/jamming phenomena \cite%
{nosso-PRE-clogging-2019}, \cite{Stock-JSTAT-2019}, although the formation
of condensates is not mandatorily observed only in counterflowing systems 
\cite{majundar-evans2005}.

Recently, we have developed a model with exclusion based on Fermi-Dirac
distribution \cite{nosso-PRE-clogging-2019} from an extension of a simple
model without exclusion previously defined in \cite%
{Roberto-pedestrian-dynamics-2015}. The easiness and coverage of this
extended model is related to the manipulation of a single parameter that can
model systems going from hard body systems with exclusion up to systems
completely random and without exclusion effects. We have previously shown
that such systems have an interesting transition from a clogging phase to a
mobile phase that we have described with many order parameters such as
mobility, Gini coefficient, over different situations \cite{Stock-JSTAT-2019}%
.

The extension of this Fermi model for the counterflowing streams of
particles in two dimensions has not yet studied and it deserves our
attention. In this paper we propose a two-dimensional version of the
Fermi-model to describe the counterflowing streams of particles on a
lattice. We show that the system can present metastable events between a
clogging and a lane phase, which to the best of our knowledge was never
observed in similar systems of particles. Thus in this paper, we
qualitatively and quantitatively explore this phenomena in details.

Our paper is organized as follows: we present one of the possible
formulations (the more appropriate in our conception) of the modeling in two
dimensions. We show that a renormalization is required for $d=2$, but that
naturally recovers $d=1$ previously studied by us in our first contributions
about this topic. After, by concluding the paper, we present our results and
a few conclusions and summaries are presented.

We start defining our scenario: a rectangular system of $V=L_{x}L_{y}$ cells
with periodic boundary conditions (toroidal lattice) where $N$ particles of
two species, namely $A$ and $B$, can move. Each cell has a maximum
occupation level denoted by $\sigma _{\max }$. Our particles are able to hop
only to its neighboring cells where particles of the species $A$ tend to
move to the $+x$ direction (along the toroid) whilst particles of the
species $B$ tend to move to the $-x$ direction, exactly as the effect of an
electric field longitudinally applied, considering that particles of species 
$A$ are oppositely charged to the particles of species $B$. In our model,
particles only interact with particles of the opposite type. Thus a particle
of species $A$ that occupies the cell $(x,y)=(i\epsilon ,j\epsilon )$ hops
to its neighboring cell $(x,y)=(i\epsilon +\epsilon ,j\epsilon )$ at the
instant $t=\tau l$ according to a Fermi-like distribution of probability
given by 
\begin{equation}
p_{(i,j)\rightarrow (i+1,j)}^{(l)}=\frac{1}{Z^{(l)}}\frac{1}{1+e^{\alpha
\left( B_{i+1,j}^{(l)}-\sigma _{\max }\right) }},
\label{Eq:prob_mov_A_front}
\end{equation}%
where $B_{i+1,j}^{(l)}$ is the number of particles of the species $B$ in the
cell, $\alpha $ is the parameter that control the randomness of the
dynamics, $\epsilon $ is the step-length (linear dimension of the cells) and 
$\tau $ is the time spend to perform the transition. Here $Z^{(l)}$ is a
normalization factor to be defined later.

We define the probability of the particles to move along the $y$-direction
as a product of two factors: scattering and diffusion. The scattering factor
is defined as the complement of the chances of a particle to move in the
direction of the toroid, i.e., the complement of the Eq. \ref%
{Eq:prob_mov_A_front}. In a similar way, the diffusion factor consist on the
chances of a particle to enter the neighboring cell in the $y$-direction,
which depends on the level of occupation of the target cell.

By this logic, we describe the chances of the particle $A$ to move to a cell
in the transversal direction of the toroid as 
\begin{equation}
p_{(i,j)\rightarrow (i,j\pm 1)}^{(l)}=\frac{e^{\alpha
(B_{i+1,j}^{(l)}-\sigma _{\max })}(1+e^{\alpha (B_{i,j\pm 1}^{(l)}-\sigma
_{\max })})^{-1}}{Z^{(l)}(1+e^{\alpha (B_{i+1,j}^{(l)}-\sigma _{\max })})}
\label{Eq:prob_mov_A_left_right}
\end{equation}%
taking in consideration the difusive and scattering factor in its
composition. Here the normalization factor is defined as: $Z^{(l)}\equiv
\sum_{\left\langle i^{\prime },j^{\prime }\right\rangle }p_{(i,j)\rightarrow
(i^{\prime },j^{\prime })}^{(l)},$ where $\left\langle i^{\prime },j^{\prime
}\right\rangle $ denotes that sum is taken over the nearest neighboring
cells to which the movement is allowed. However, to keep the similarities
with the original model, the particles also have to have the chances of
remaining in its cells. So we also define the probability of a particle $A$
to stay in its cell as the complement of the sum of all probabilities of
movement. But that part is tricky, because we must take into account that $%
0\leq Z^{(l)}\leq 2$ and thus we have to make use of a necessary simple
algorithm in our MC simulations to overcome the two following possible
situations:

\begin{itemize}
\item $Z^{(l)}\leq 1$: in this scenario we have $p_{(i,j)\rightarrow
(i,j)}^{(l)}\equiv 1-Z^{(l)}$ and the equations \ref{Eq:prob_mov_A_front},
and \ref{Eq:prob_mov_A_left_right}, are not normalized by the factor $%
Z^{(l)} $;

\item $Z^{(l)}>1$: in this case the particle will certainly move and the
equations \ref{Eq:prob_mov_A_front}, and \ref{Eq:prob_mov_A_left_right} are
normalized by the factor $Z^{(l)}$.
\end{itemize}

By symmetry, a particle of the species $B$ in the cell $(x,y)=\epsilon (i,j)$
hops to its neighboring cell $(x,y)=(i\epsilon -\epsilon ,j\epsilon )$ at
the instant $t=\tau l$ with probability: $p_{(i,j)\rightarrow
(i-1,j)}^{(l)}=(1/Z^{(l)})(1+e^{\alpha \left( A_{i-1,j}^{l}-\sigma _{\max
}\right) })^{-1}$ and similarly: $p_{(i,j)\rightarrow (i,j\pm
1)}^{(l)}=(1/Z^{(l)})e^{\alpha (A_{i-1,j}^{(l)}-\sigma _{\max
})}(1+e^{\alpha (A_{i,j\pm 1}^{(l)}-\sigma _{\max })})^{-1}/(1+e^{\alpha
(A_{i-1,j}^{(l)}-\sigma _{\max })})$.\qquad

The main parameter of the adapted Fermi-like distribution in our model is $%
\alpha $ once it controls the randomness of the dynamics that can range from
a purely random scenario to a deterministic case where the relation between
the cell's occupation and the limitation factor ($\sigma _{\max }$) is
paramount. In the context of counter-flowing particle dynamics, $\alpha $
can be associated with the magnitude of an external field applied to
oppositely charged particles.

We can understand the influence of $\alpha $ on the dynamics by focusing on
the two extreme situations: (i) $\alpha \rightarrow 0$ and (ii) $\alpha
\rightarrow \infty $. In the case (i) ($\alpha \rightarrow 0$) the dynamics
is completely random with particles $A(B)$ moving cell by cell with
probability $1/2$ towards the $+(-)x$ direction if the particle is from the $%
A(B)$ species and with $1/4$ to the $\pm y$ direction. It is easy to notice
that there is no chance of particles to stay in its cells. On the other
hand, in the contrasting situation (ii) ($\alpha \rightarrow \infty $)
however, the system behaves with deterministic dynamics such as the lattice
gases and the set of possible configurations of occupation of the
surrounding cells will define with certainty the movement of all particles
except for the particular case in which the target cell happen to have $%
\sigma =\sigma _{\max }$. In that case, a local random decision is made as a
result of $p=1/2$ or $1/4$ depending on the direction of the occupied cell,
although we do not expect it to substantially alter the bulk behavior of the
system. Moreover, it is important to mention that we have a notorious
computational advantage in using this model, since we are not obliged in
checking by brutal force if \ cells satisfies the occupation limit $\sigma
_{\max }$ such as systems with nearest neighbor exclusion \cite%
{dickman-lattice-gas1988,
dickman-lattice-gas-first-second-order-transition2001}.

To describe the state of the dynamics we define the mobility, denoted as $%
M(t)=\frac{1}{N}\sum_{i}^{N}\xi _{i}(t)$, where $\xi _{i}(t)$ is a binary
parameter that assumes the value $1$ when a particle moves in the $x$%
-direction and assumes the value $0$ otherwise. The mobility stands as a
current of particles or simply the fraction of the particles that perform a
step along its specie's drift direction. When jamming/clog occurs the
mobility drops to zero. While when an opposite scenario occurs like
particle's self-organization in lane formation, the mobility will tend to
one as a result of the optimal flow. To capture the lane-organized state we
define the order parameter in the $x$-direction, or simply \textquotedblleft 
$x$-order parameter\textquotedblright , denoted as $\Phi _{x}^{l}\equiv 
\frac{1}{N}\sum_{k=1}^{Ly}|A_{k}^{l}-B_{k}^{l}|$, where $A_{k}^{l}\equiv
\sum_{j=1}^{Lx}A_{j,k}^{l}$ and $B_{k}^{l}\equiv \sum_{j=1}^{Lx}B_{j,k}^{l}$
are the marginal sum over the $j$ variable (cells along the $y$-direction)
of the number of particles of each species. By the definition of $\Phi
_{x}^{l}$ we see that it will tend to one when the system is in a
lane-organized state along the toroid which represents an optimal state of
flow due to the lack of particles moving in the opposite direction.

The mobility parameter is only possible to be calculated when performing MC
simulations, differently from $\Phi _{x}$ that depends only on the spacial
concentration of particles and can also be measured when one integrates the
recurrence relations that emerge from a mead field approximation of the Eqs. %
\ref{Eq:prob_mov_A_front}, and \ref{Eq:prob_mov_A_left_right} for the
species $A$ as well as it can be done for the species $B$, similarly to what
was made in Ref. \cite{mestrado-eduardo-roberto2017}.

We implemented standard MC simulations where the MC step consists of
choosing $N$ particles at random of the $N$ possible particles and
sequentially updating its position before moving to the next MC step. In all
of our results we use $N_{A}=N_{B}=N/2$ and a number of runs following the
relation $N_{run}=100000/N+10000$ unless told otherwise.

A detailed study of the influence of $\alpha $\ and $\sigma _{\max }$\ were
explored in our previous contributions in one dimensional systems \cite%
{nosso-PRE-clogging-2019}, \cite{Stock-JSTAT-2019}, and in this current
paper, we concentrate our study in a situation where the clogging is
expected ($\alpha $ large and $\sigma _{\max }=1$) and therefore the most
important situation to be explored. \textbf{\ }

\begin{figure}[th]
\begin{center}
\includegraphics[width=1.0\columnwidth]{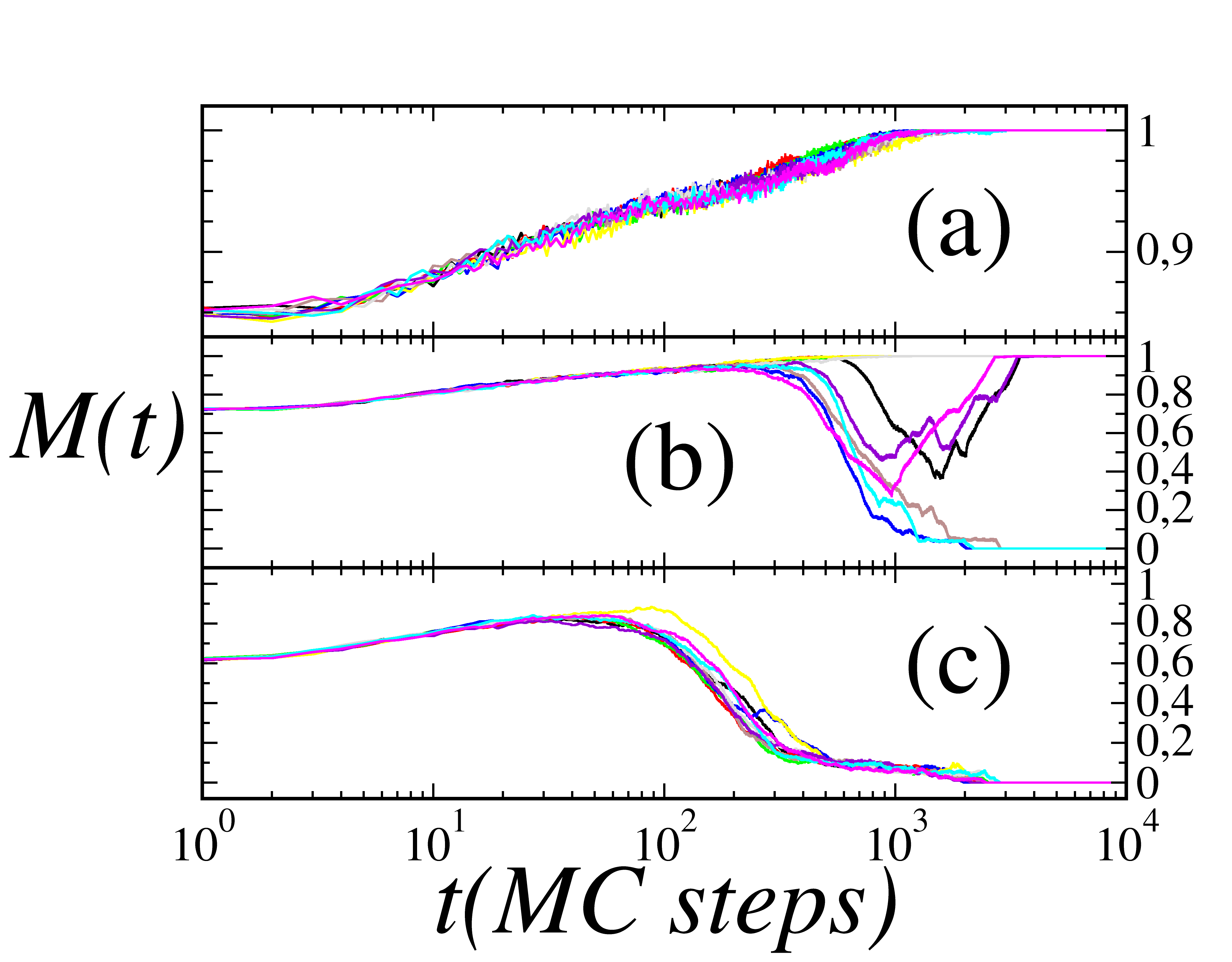}
\end{center}
\caption{Time evolution of mobility for ten different runs (different seeds)
of a system with dimensions of $L_{x}=256$ and $L_{y}=64$ and with $\protect%
\alpha =20$. The dynamics relaxes to a state of optimal flow when $\protect%
\rho =0.5$ (a) and relax to \textcolor{red}{a}/\textcolor{blue}{an} immobile
state when $\protect\rho =1.5$ (c). For $\protect\rho =1.0$ (b), the system
presents a metastable steady state where either the mobile or the immobile
phase can arise depending sensibly on the initial conditions.}
\label{Fig:mob_evo_1}
\end{figure}

\begin{figure}[th]
\begin{center}
\includegraphics[width=1.0\columnwidth]{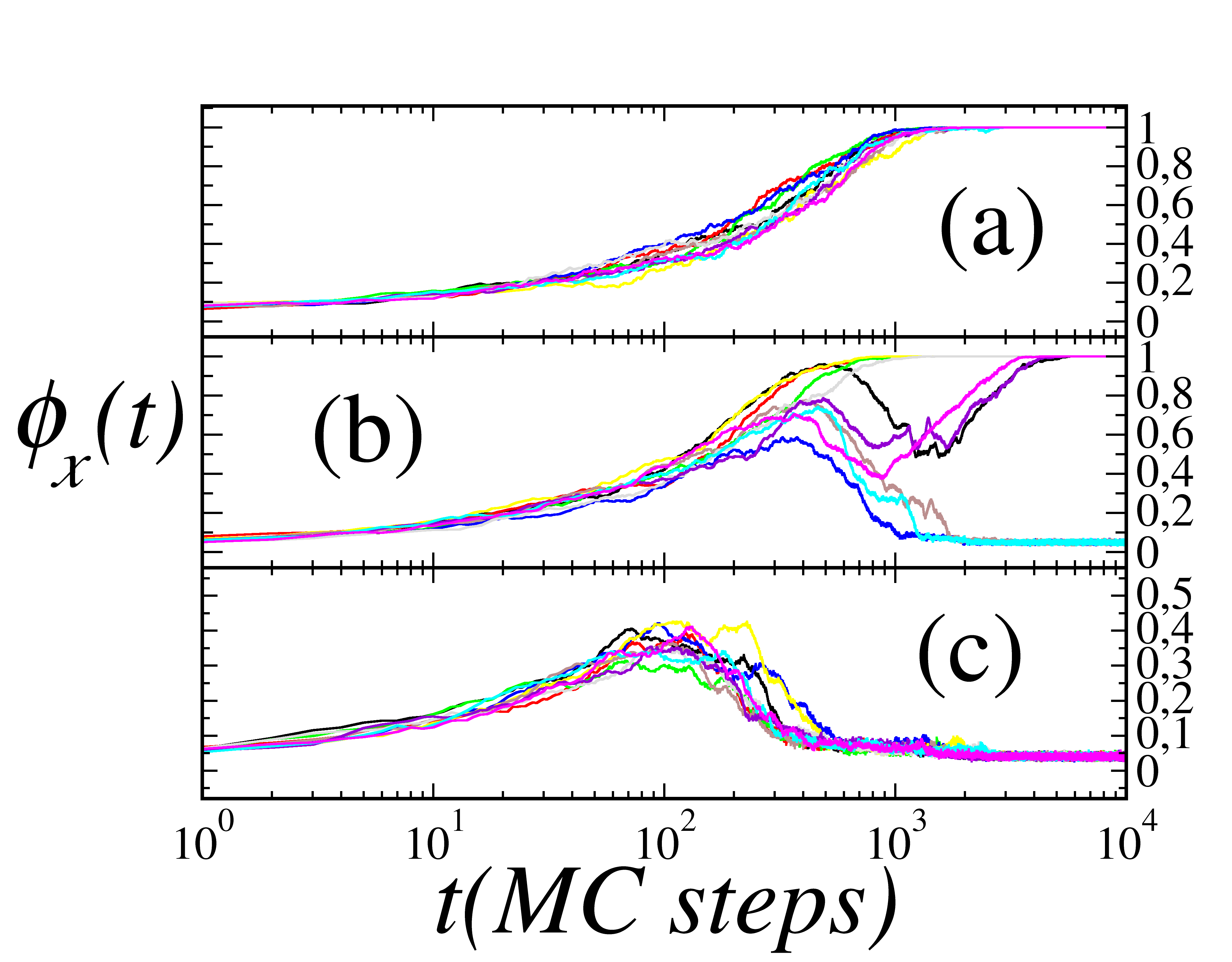}
\end{center}
\caption{Time evolution of the alternative order parameter $\Phi _{x}$ under
the same parameters used in the previous figure \protect\ref{Fig:mob_evo_1}
for the mobility. Here the conclusions are exactly the same that one obtains
for the time evolution of the mobility.}
\label{Fig:xop_evo_1}
\end{figure}

Our first result is to analyze the time evolution of mobility and $\Phi _{x}$
for considerably low stochastic level $\alpha =20\ $\cite{footnote1} and for
three different values of densities: $\rho =0.5$, $\rho =1.0$ and $\rho =1.5$%
. At this regime we expect the system of dimensions $L_{x}=256$ and $L_{y}=64
$, to present a high mobility scenario for low density whilst for high
density case we expect the system to jam. The Figs. \ref{Fig:mob_evo_1} and %
\ref{Fig:xop_evo_1} shows the expected behavior for the two present extreme
cases of density (plot (a) and (c)) as we can see $M(t)$ and $\Phi _{x}(t)$
for each of the ten runs rapidly reaching the steady state. The case of
intermediate density, i.e., $\rho =1.0$, surprisingly shows a metastable
steady state as we can see by the plot (c) in the Figs. \ref{Fig:mob_evo_1}
and \ref{Fig:xop_evo_1}. The present case reveals that for a certain
specific density, the steady state of each run depends sensibly on the
initial conditions as we can observe by the splitting of the mobility and
the $x$-order parameter. We can also observe in the plot (c) of Figs. \ref%
{Fig:mob_evo_1} and \ref{Fig:xop_evo_1} that it is a bimodal situation,
since there is no intermediate steady state of mobility other than the fully
mobile self-organized state or the jammed state. 

To better understand this metastability phenomena, we studied the system's
dependence on the density $\rho =N/V$. By knowing that the dynamics relax to
a bimodal steady state of extremely different outcomes, where for each run
the mobility and $\Phi _{x}$ do not change after a sufficient time, we
implemented the same criteria to stop the time evolution as used in the
previous work based on the slope of the time dependent curve to consider
when the system has reached the steady state. The considered criteria
consists on making a linear fitting on $M(t)$ and $\Phi _{x}(t)$ averaged
over consecutive large number of time intervals $\Delta t$ and so comparing
the slope of the fitted curve to an error value $\eta $. In this work, we
used $\Delta t=1000$ and $\eta =10^{-7}$ as it was shown in Ref. \cite%
{nosso-PRE-clogging-2019} to be a good choice. Now focusing on the study of
the density's influence on the metastability behavior, we simulated a total
of $N_{run}=10^{5}/N+10^{4}$ different runs to calculate the probability of
lane formation (the number of times that the system reaches the lane pattern
at steady state divided by $N_{run}$), which is denoted as $p_{lane}$,
similarly the probability of clog formation, denoted as $p_{clog}$, and the
probability of any outcome other than the two previous cases, denoted as $%
p_{else}=1-p_{clog}-p_{lane}$.

\begin{figure}[bh]
\begin{center}
\includegraphics[width=1.0\columnwidth]{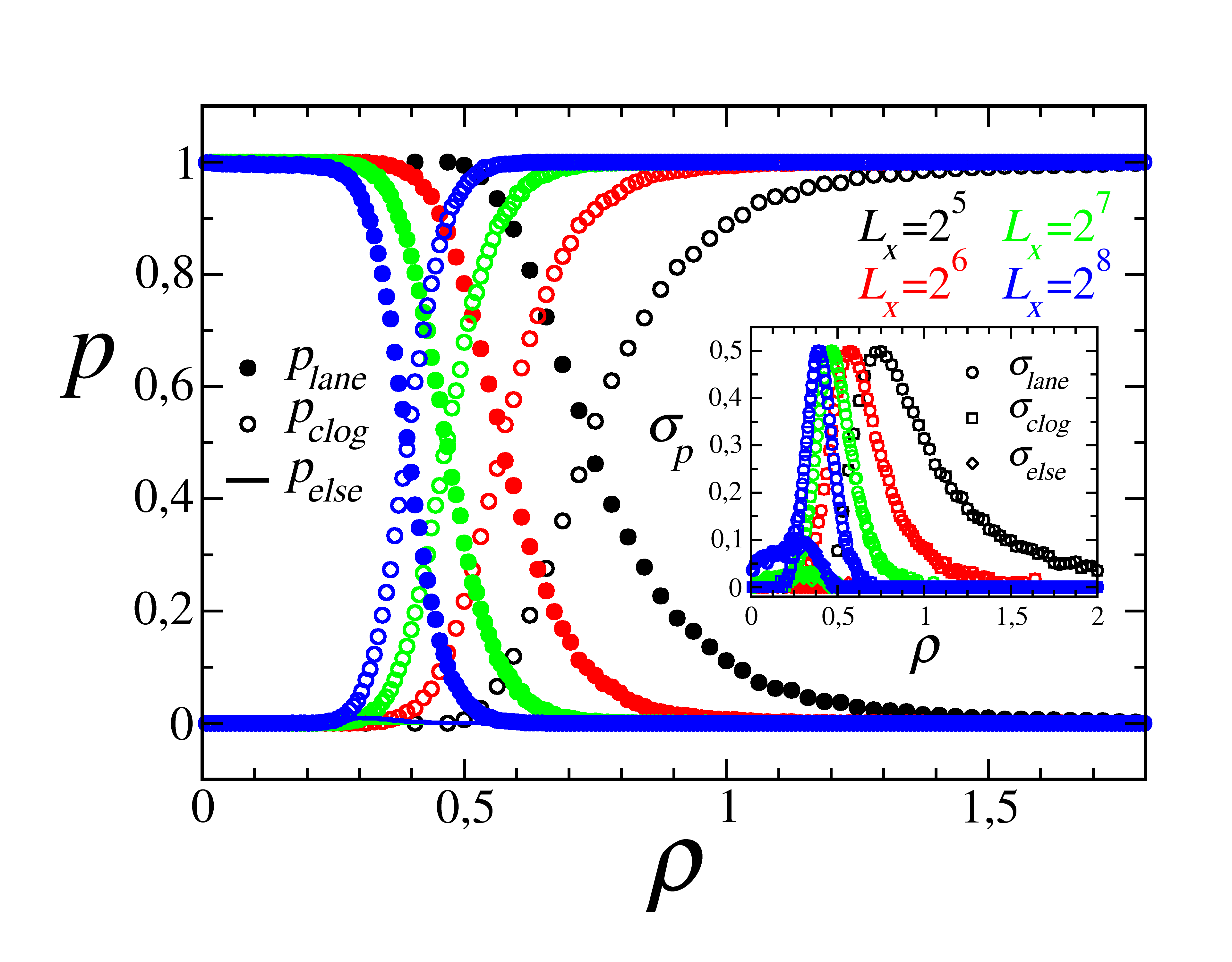}
\end{center}
\caption{The probabilities of lane formation (filled circles), jamming
(empty circles) and relaxation to any other state in between (lines) for a
system with width of $L_{y}=2$ for $L_{x}=2^{5}$ (black), $L_{x}=2^{6}$
(red), $L_{x}=2^{7}$ (green) and $L_{x}=2^{8}$ (blue). We are able to see
that the crossover density asymptotically decreases as the length of the
system increases. The inset plot shows the standard deviation of the
probabilities having their maximum value $\protect\rho _{c}$ decreasing with
lengthy systems.}
\label{Fig:probs_fss_length}
\end{figure}

The Fig. \ref{Fig:probs_fss_length} shows us the probabilities the system
have of reaching each one of the three possible steady states as a function
of $\rho $ for a system of width $L_{y}=2$ (purposefully quite narrow) and
different values of length $L_{x}=2^{5}$ (black), $L_{x}=2^{6}$ (red), $%
L_{x}=2^{7}$ (green), and $L_{x}=2^{8}$ (blue). We can observe that the
system shows the expected lane formation state for small densities
regardless of the size of the system. When the system has sufficiently large
density ($\rho \geq 0.5$ for $L_{x}=2^{5}$ for example) clogging starts to
emerge as stationary state and the bimodal meta-stable scenario arises. With
larger density values, $p_{clog}$ keeps increasing whilst $p_{lane}$
decreases until both curves crosses each other over and the most probable
scenario for the dynamics shifts.

The inset plot of Fig. \ref{Fig:probs_fss_length} shows us that the standard
deviation of the $p$ has its maximum value at the crossover density where $%
p_{lane}=p_{clog}=1/2$ and thus we define the crossover density $\rho
_{c}\equiv \rho (p_{lane}=p_{clog}=1/2)$. As the Fig. \ref%
{Fig:probs_fss_length} suggests, the lengthier the system is, the more the
crossover density appears to decrease asymptotically and at the same time
the range of density at which the metastability phenomena occurs becomes
smaller with larger system. 

\begin{figure}[th]
\begin{center}
\includegraphics[width=1.0\columnwidth]{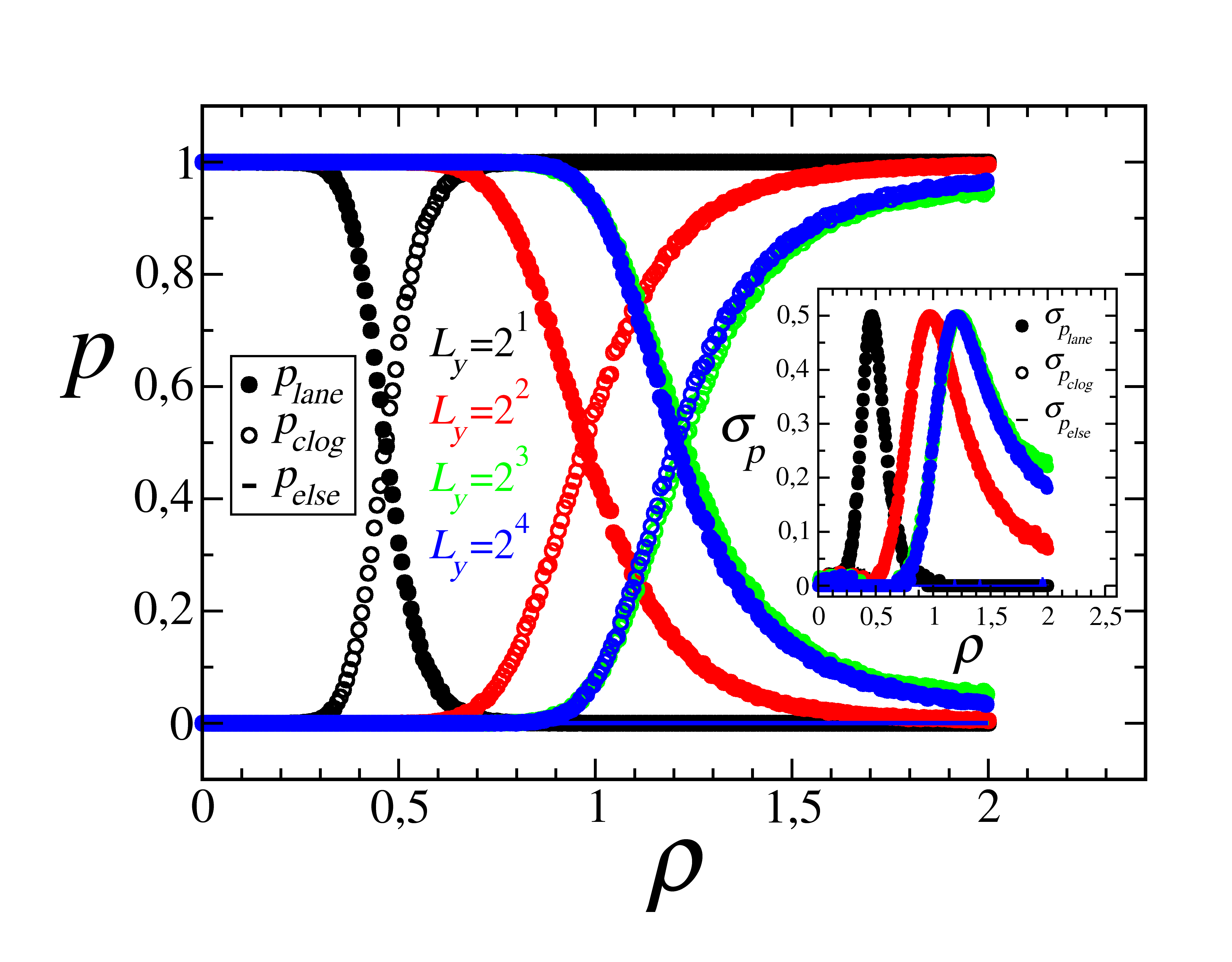}
\end{center}
\caption{The probabilities of lane formation (filled circles), jamming
(empty circles) and relaxation to any other state (lines) for a system of
length $L_{x}=128$. We plotted the results for different widths: $L_{y}=2$
(black), $L_{y}=2^{2}$ (red), $L_{y}=2^{3}$ (green), and $L_{y}=2^{4}$
(blue). We notice that the crossover density increases from $\protect\rho %
_{c}\approx 0.5$ up to a maximum $\protect\rho _{c}\approx 1.2$ for systems
with $L_{y}\geq 2$.}
\label{Fig:probs_fss_width}
\end{figure}

Let us continue with our finite size scaling study, but now one changes the
width of a long system. We show in the Fig. \ref{Fig:probs_fss_width} the
probabilities of the stationary outcome for different widths $L_{y}=2$
(black), $L_{y}=2^{2}$ (red), $L_{y}=2^{3}$ (green), and $L_{y}=2^{2}$
(blue) whilst keeping the length fixed on $L_{x}=128$. We observe that when
the toroid is considerably narrow the system presents a great difference on
the $\rho _{c}$ as we can see it increases for the smaller widths $L_{y}=2$, 
$L_{y}=2^{2}$, and $L_{y}=2^{3}$. However, when the system has a width of $%
L=2^{3}$ or greater, the crossover density does not change and we would
expect the chances of lane formation to increase with wider corridors but it
does not occur due to the proportionally larger number of particles.

\begin{figure}[bh!]
\begin{center}
\includegraphics[width=1.0%
\columnwidth]{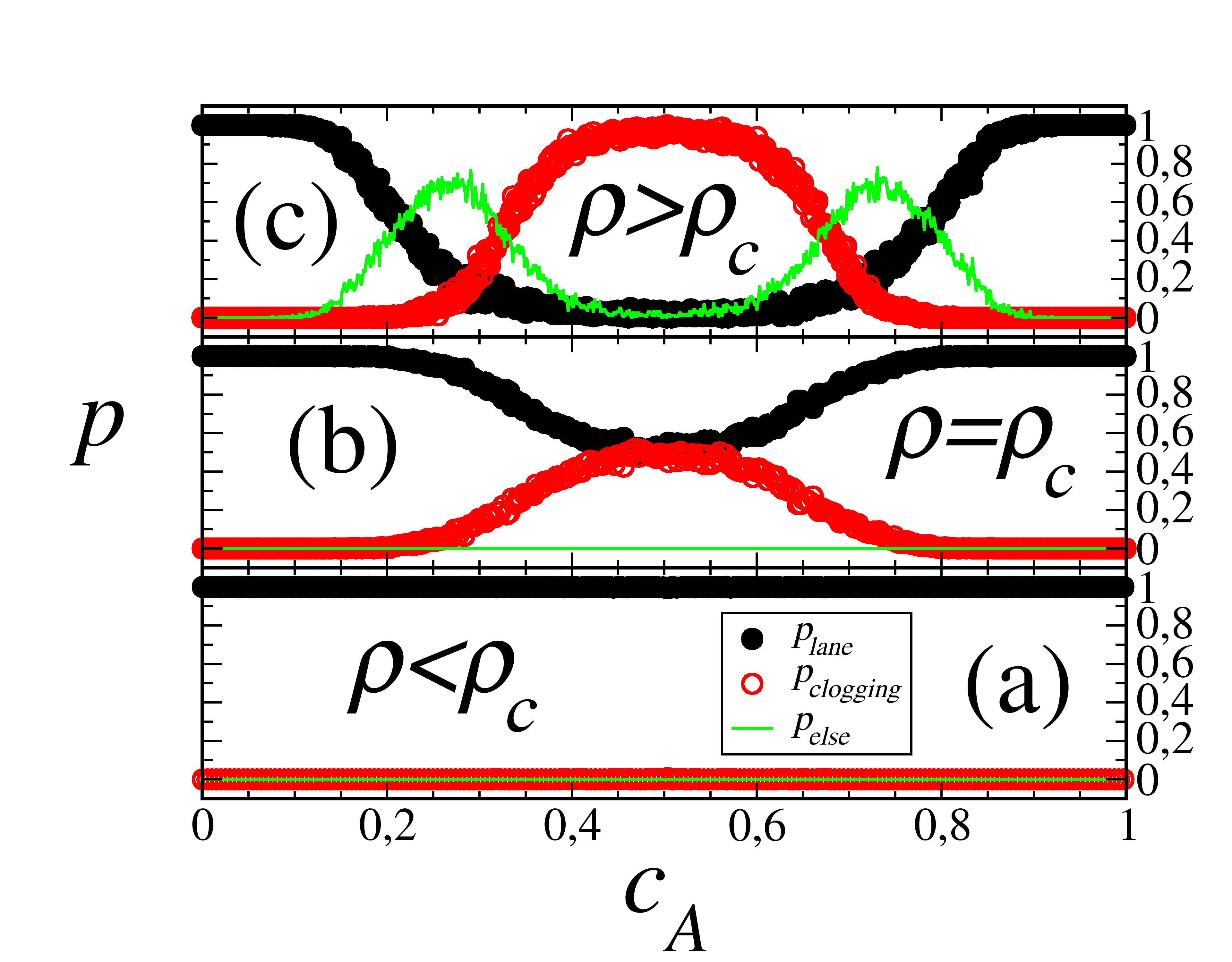}
\end{center}
\caption{The probabilities ($p_{lane}$, $p_{clog}$, and $p_{else}$)
dependence on the relative density for three values of the absolute density.
For $\protect\rho <\protect\rho _{c}$ (plot a) the system will flow
regardless of the mixing of species. For $\protect\rho =\protect\rho _{c}$
(plot b) the system has $p_{lane}=1$ for a $c_{A}\leq 0.1$ and falls to $%
p_{lane}=0.5$ when the mixing is about 1:1 as is defined the crossover
density. Finally when $\protect\rho >\protect\rho _{c}$ (plot c) the system
presents an intermediate behavior of non-zero mobility without lane
formation in the interval of $0.1<c<0.4$.}
\label{Fig:probs_relative_density}
\end{figure}

We also approached the problem considering different relative concentration
of particles, i.e. different mixing of species. The Fig. \ref%
{Fig:probs_relative_density} shows the stationary state of the dynamics when
the proportion of the species ranges from a scenario of basically one
species moving along the toroid ($c_{A}=0$, where $c_{A}+c_{B}=\frac{%
N_{A}+N_{B}}{N}=1$) to the case studied until this point of equal
concentrations $c_{A}=c_{B}=1/2$. In the present plot (Fig. \ref%
{Fig:probs_relative_density}) we show the three regimes that arises when $%
\rho <\rho _{c}$, $\rho =\rho _{c}$, and $\rho >\rho _{c}$. For a system
with dimensions $L_{x}=128$ and $L_{y}=16$ the crossover density is $\rho
_{c}\approx 1.2$ (blue curve in the Fig. \ref{Fig:probs_fss_width}), we
therefore simulated the cases $\rho =0.5$, $\rho =1.2$, and $\rho =3.0$
which are shown in the plots of Fig. \ref{Fig:probs_relative_density} (a),
(b), and (c) respectively.

For all the curves we begun by simulating a scenario of one species ($%
c_{A}=0 $), which means $p_{lane}=1$ and obviously it represents a fake
state of lane formation for the fact that there is not induced
self-organization when only one species is considered.

Fig. \ref{Fig:probs_relative_density} (a) shows an expected behavior for the
dynamics because the level of occupation ($\rho $) of the system assures the
lane formation despite the growth of the concentration of the species $A$.
However, the case of $\rho =\rho _{c}$ ( Fig. \ref%
{Fig:probs_relative_density} (b) ) shows that $p_{lane}$ gradually decreases
as $c_{A}$ grows. Finally in Fig. \ref{Fig:probs_relative_density} (c) which
corresponds $\rho >\rho _{c}$, we observe the arise of a new possibility: a
transitional scenario where the system do not self-organize itself and
neither relax to a jammed state. In that case, the overall density $\rho $,
is above the crossover level so the growth of $c_{A}$ around $c_{A}\approx
0.2$ makes the particles $A$ stand as impurities for the movement of $B$ but
are not sufficient to make system immobile. In this case we can have self
organized systems and also unorganized motion patterns, which makes us
believe that the term \textquotedblleft chaotic\textquotedblright\ is the
best suited to describe it since the system can lose its ordering pattern.
When the mixing of species tend to the same value ($c_{A}\approx c_{B}$) the
system essentially shows the clogging steady state uniquely.

\begin{figure}[bh!]
\begin{center}
\includegraphics[width=1.0%
\columnwidth]{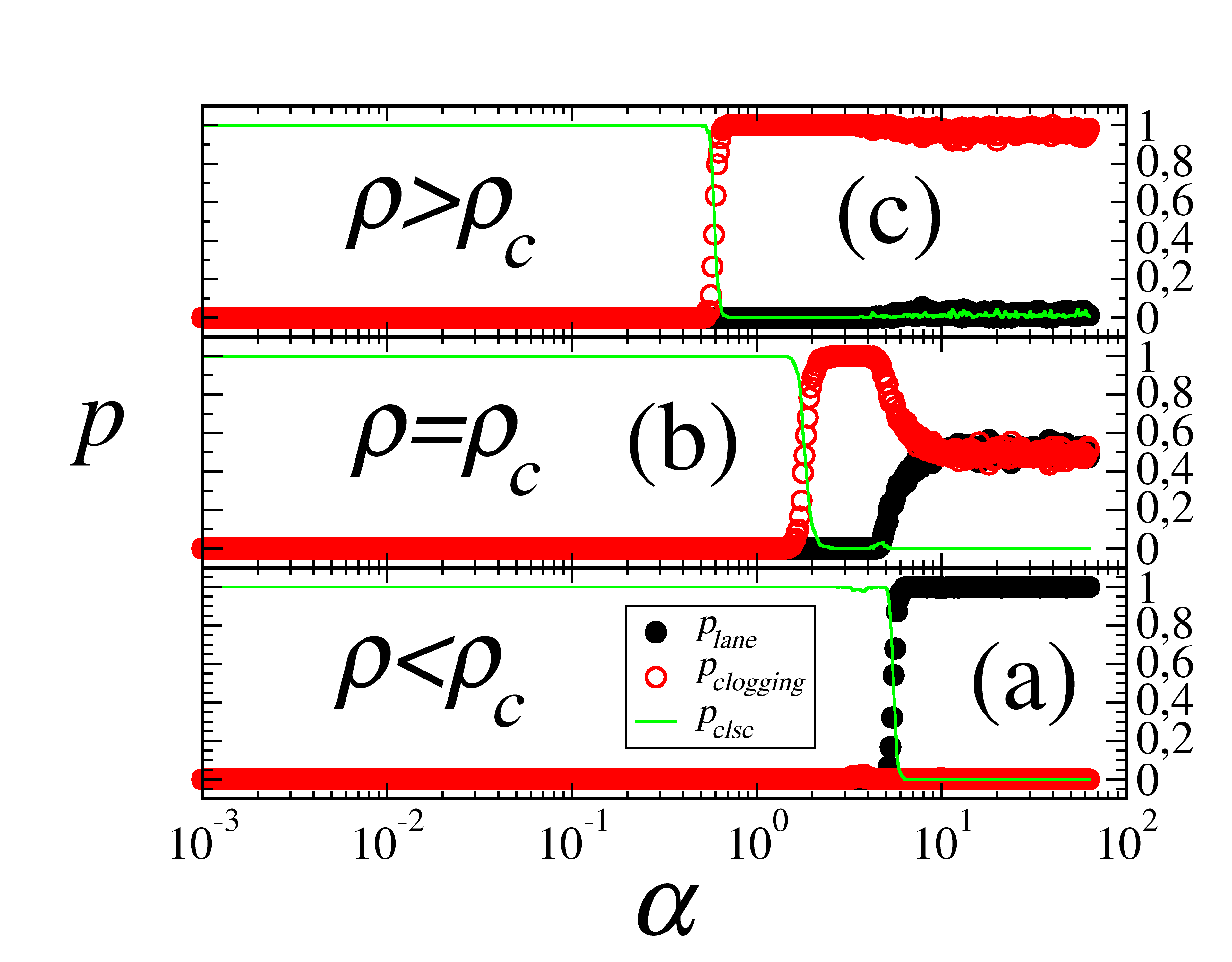}
\end{center}
\caption{Behavior of $p$ as function of $\protect\alpha $ considering $%
c_{A}=c_{B}$, for the same size and other parameters used in Fig. \protect
\ref{Fig:probs_relative_density}. The metastability occurs only for $\protect%
\alpha \gtrsim 10$ and $\protect\rho =\protect\rho _{c}$. }
\label{Fig:alfa_value}
\end{figure}

In order to conclude, it is important to comment some points about $\alpha $%
. We used a sufficiently large value of $\alpha $ during all our manuscript.
But this was not duly justified. Here we performed simulations to obtain a
plot of $p\times \alpha $ considering the same size system ($L_{x}=128$ and $%
L_{y}=16$), keeping $c_{A}=c_{B}$ which was used for the most part of this
manuscript, and using $\rho =0.5<\rho _{c}$ (plot (a)), $\rho =\rho _{c}=1.2$
(plot (b)), and $\rho =3.0>\rho _{c}$ (plot (c)) in Fig. \ref{Fig:alfa_value}%
.

The result suggests that only for $\rho =\rho _{c}$, and $\alpha \gtrsim 10$
we can observe the metastability, but for small values of $\alpha $, the
system is moderately mobile which is expected since that the interaction
between opposite species is weak.

By concluding, in this paper we extend for two dimensions the Fermi-like
model for the counterflowing streams of particles. We show that for a
suitable density of particles the system under balance/symmetry between the
species: $c_{A}=c_{B}$, the system is metastable since it evolves to only
two possible steady states: one characterized for a immobile regime
(clogged) and another mobile self-organized regime represented by a pattern
of lanes of particles. Our results also suggest that an imbalance/asymmetry
between the species: $c_{A}\neq c_{B}$, can generate in addition to these
two patterns, an alternative situation characterized by a random mobile
state since the dominance of one of species, makes the system to
macroscopically behave as a single flux.

\bigskip


\begin{thebibliography}{99}
\bibitem{footnote1} For $\alpha =20$ the system is essentially
deterministic, which is exactly our purpose since only in this regime the
metastable states can ocurr. For this reason, we concentrate a discussion
about $\alpha $ in the final of this paper presenting a plot to justify this
initial choice of our study.
\end{thebibliography}
\end{document}